\documentclass[subscriptcorrection,upint,varvw,mathalfa=cal=euler,balance,hyphenate,french,pdf-a,nolists]{asmejour} %

\usepackage{ulem}
\usepackage{xcolor}

\DeclareRobustCommand{\Erase}{\bgroup\markoverwith{\textcolor{red}{\rule[.5ex]{2pt}{1.4pt}}}\ULon}

\hypersetup{%
	pdfauthor={John H. Lienhard},                       		   	
	pdftitle={ASME Journal Paper LaTeX Template},                  	
	pdfkeywords={ASME journal paper, LaTeX template, BibTeX style, asmejour class},
	pdfsubject = {Describes the asmejour LaTeX template},			
	pdflicenseurl={https://ctan.org/pkg/asmejour},
}

\JourName{Applied Mechanics}

\begin{document}


\SetAuthorBlock{Kotone Tajiri}{
   School of Integrated Design Engineering, Graduate School of Science and Technology,\\
   Keio University,\\
   3-14-1 Hiyoshi,\\
   Kohoku-ku, Yokohama, Japan \\
   email: kotone_tajiri@keio.jp}

\SetAuthorBlock{Tomohiko G. Sano\CorrespondingAuthor}{
   Department of Mechanical Engineering,\\
   Keio University,\\
   3-14-1 Hiyoshi,\\
   Kohoku-ku, Yokohama, Japan \\
   email: sano@mech.keio.ac.jp}


\title{Buckling of knitted fabric wrapped around a rigid cylinder}

\keywords{Knitted fabric, Buckling, Wrinkling}

\begin{abstract}
Knitted fabrics exhibit high flexibility due to their periodic loop structures formed by bent yarns. Under compressive loading, they develop three-dimensional (3D) wrinkling patterns that reflect nonlinear interactions between yarn elasticity and local loop deformations, as observed when the sleeves of a sweater are rolled up. Despite their widespread use in garments and medical textiles, the relationship between loop-level geometry and macroscopic buckling remains less understood.
Here, we investigate the 3D deformation of knitted fabrics wrapped around a rigid cylinder under uniaxial compression. Circumferential and axial stitch numbers are systematically varied to determine how loop geometry affects the evolution of wrinkle patterns. Samples with a small number of circumferential stitches exhibit sequential wrinkle formation from the compressed end, leading to an accordion-like wrinkle pattern, whereas those with a larger number of stitches form helical wrinkles simultaneously across the surface. Wrinkle morphology changes progressively with stitch geometry, accompanied by systematic variations in compressive force, loop deformation, and helical wrinkle angle. The development of helical wrinkles originates from subtle structural asymmetries introduced during manufacturing processes, including the tension applied during knitting and the direction of sample assembly. 
These results demonstrate that small variations in local loop deformation can lead to substantial differences in wrinkle morphology, highlighting the sensitivity of macroscopic buckling to microscopic structural features. The study establishes a direct link between loop-level mechanics and global deformation behavior, providing a basis for the predictive design of knitted structures with tailored mechanical responses and complex 3D patterns.
\end{abstract}

\date{\today}

\maketitle 


\section{Introduction}

When the sleeves of a sweater are rolled up, complex wrinkle patterns form near the cuffs. Surface instabilities of compliant thin films under compression are ubiquitous in natural systems and increasingly exploited in engineering applications \cite{Chen2010, Li2012}. Biological examples include wrinkles formed during eyelid motion and folding structures in turtle skin \cite{Zhu2013}, as well as morphogenesis-related patterns in the brain \cite{David1975, Van1997}, intestine \cite{Ben2012, Ciarletta2012}, and kidney \cite{Satu2000}. Buckling instabilities have furthermore been utilized for micro- and nanoscale patterning \cite{Romain2015}, microfluidic systems \cite{Efimenko2005}, and defect engineering in elastic surface crystals \cite{Francisco2016}. Most prior studies have examined stiff films bonded to compliant substrates, leading to periodic wrinkles on flat \cite{Cai2011} and curved \cite{Xu2016, Xu2017} surfaces. Excessive stress can induce symmetry breaking, generating localized \cite{Jin2015} or disordered patterns \cite{Li2011}.
Other examples of model systems that exhibit wrinkling include rucks in a rug~\cite{Vella2009} and films on cylindrical substrates~\cite{Yang2018}.
Compared with flat systems, curvature enriches the instability modes by fundamentally altering the energetic landscape \cite{Stoop2015}.

Knitted fabrics form a highly deformable class of materials whose mechanical responses originate from networks of interlaced loops. In a plain (stockinette) knit, a periodically repeated loop topology in the vertical and horizontal directions yields an architecture of bent, interlocking yarn segments \cite{Amanatides2022}. This looped geometry grants knitted textiles exceptional geometric compliance, and individual loops rotate, slide, and reconfigure under loading. As a result, knitted materials exhibit large and reversible deformations that do not occur in continuous elastic sheets.
Given this microstructural origin of deformation, understanding how loop-level mechanics govern macroscopic responses is a central challenge in textile mechanics. Classical studies modeled knitted loops using geometric and force-based representations \cite{Chamberlain1926, Leaf1955, Hearle1969}. More recent approaches incorporate yarn dynamics, frictional contact, and finite-element simulations to capture loop rearrangement, anisotropy, and nonlinear tensile behavior \cite{Kaldor2008, Cirio2016, Cherradi2022, Minapoor2015}. These efforts demonstrate that the effective moduli, anisotropies, and nonlinear characteristics of knitted textiles originate from the topology of the loop network. Our previous study demonstrates that the interplay between yarn curvature and contact constraints significantly affects overall fabric compliance~\cite{Kotone2025}.

Buckling in knitted fabrics, such as the wrinkles that appear when sleeves are rolled up, arises from nonlinear interactions within the looped architecture rather than from simple elastic bending. Understanding such instabilities is essential for improving garment performance and also for establishing design principles for advanced textile structures. However, the role of deformation within individual knitted unit cells in determining macroscopic wrinkle morphology remains unclear. The connection between loop-level kinematics and large-scale buckling patterns remains insufficiently established.
This gap is especially pronounced for compressive loading, which induces inherently three-dimensional instabilities not captured by conventional two-dimensional or tension-dominated models \cite{Poincloux2018}. Clarifying this relationship is essential for developing mechanistic descriptions of instability in knitted materials and for enabling the rational design of fabrics with tailored mechanical and functional properties.

\begin{figure*}[h!]
    \centering\includegraphics[width=0.9\linewidth]{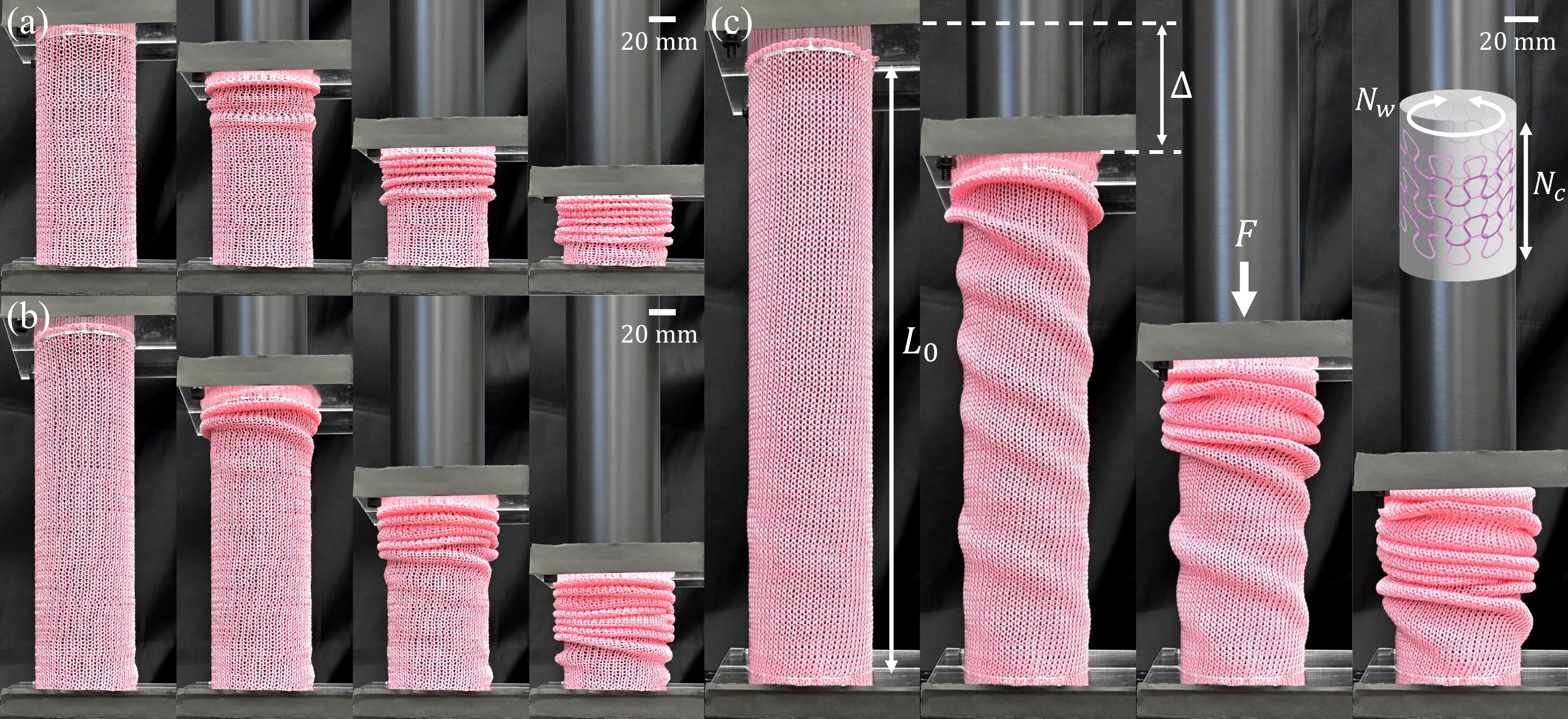}
    \caption{Snapshots of the compression process for knitted samples wrapped around a rigid cylinder. The sets of walewise (circumferential) and coursewise (axial) knitting numbers, $(N_w, N_c)$, are (a)~(40, 80), (b)~(50, 100) and (c)~(60, 120). 
    Each sample is compressed to 30\% of its natural length $L_0$ using an acrylic plate mounted on the force-testing machine. Snapshots of compression levels of 0\%, 20\%, 50\%, and 70\% are shown here. We apply a displacement, $\Delta$, to measure the compression force, $F$, and analyze wrinkles on the knit surface.
    }
    \label{fig:1}
\end{figure*}

Knitted materials have therefore attracted growing interest across apparel, medical, and engineering contexts \cite{Paradiso2005, Barhoumi2018, Granberry2019, Luo2022}. Their flexibility, stretchability, and capacity for large, reversible deformations originate from periodic loop architectures formed by interlaced yarns. Unlike woven textiles, whose mechanical responses are governed primarily by yarn stiffness and inter-yarn friction, knitted systems deform largely through geometry-driven mechanisms, which enable tunable compliance and complex shape changes under loading \cite{Crassous2024}. This structural programmability positions knitted fabrics as promising candidates for deformable interfaces, wearable devices, and morphable structures with engineered mechanical functions \cite{Nguyen2020, Tanaka2024, Pasquier2024, Pasquier2025}.
Recent advances in architected soft materials further motivate efforts to establish predictive links between yarn-scale properties, loop geometry, and emergent large-scale mechanical responses. Studies have highlighted how microscopic features such as loop curvature, bending stiffness, and contact constraints govern deformation behaviors that remain difficult to capture using continuum theories \cite{Poincloux2018, Crassous2024}. Establishing predictive relationships between loop-level mechanics and emergent large-scale deformation remains a fundamental challenge.

Textile materials composed of flexible yarn networks can undergo large, spontaneous shape changes when released from tension or subjected to external loading. A familiar manifestation is the twisting and distortion observed in lightweight knitted garments. In plain knitted fabrics, this behavior appears as \textit{spirality}, a characteristic twisting deformation in which the fabric rotates into a helical shape after production or relaxation \cite{Pavko2015}. \textit{Spirality} originates from structural asymmetries introduced during knitting, including loop geometry, yarn twist direction, and knitting direction, each of which is influenced by machine parameters and the orientation of fabric formation \cite{Raichurkar2011, Shahid2011, Shahid2015, Kalkanci2019}. 
Experimental observations indicate that machine-induced residual torque is a dominant contributor to \textit{spirality}, and that loop asymmetry, yarn twist, and knitting direction jointly determine both the direction and magnitude of the deformation \cite{Pavko2015}. Although \textit{spirality} is often regarded as a defect because it induces dimensional distortion, it also exemplifies the strong coupling between yarn-level asymmetry and macroscopic shape changes.
When curvature is introduced, such as when fabrics are wrapped into cylindrical geometries, these coupling effects are expected to interact with geometric confinement. 
When knitted fabrics are wrapped into cylindrical geometries, curvature introduces geometric confinement that interacts with the intrinsic periodicity of the loop network, giving rise to deformation modes not observed in planar configurations. Cylindrical confinement promotes complex three-dimensional responses, including twisting, torsional bias, and various forms of surface modulation associated with elastic instabilities. 
This motivates a systematic investigation of the formation of instability in cylindrical knitted architectures.

In this study, we experimentally investigate how knitted fabrics conforming to cylindrical geometries deform in three dimensions under axial compression ~{(Fig.~\ref{fig:1})}.
In plain knitted fabrics, the columns of loops oriented vertically are referred to as wales, while the rows of loops oriented horizontally are referred to as courses~\cite{Amanatides2022, Kotone2025}.
{Hereafter, walewise denotes the orientation along which wales are arranged, whereas coursewise denotes the orientation along which courses are arranged.
When a fabric is wrapped into a cylindrical geometry, these orientations naturally map onto the circumferential (walewise) and axial (coursewise) orientations of the cylinder.}
By controlling the number of stitches arranged along the circumference and length of the cylinder—which correspond, respectively, to the walewise (circumferential) and coursewise (axial) knitting numbers—we systematically assess how loop architecture influences the onset and progression of surface instabilities. Our experiments reveal that knits wrapping tightly around the cylinder develop wrinkles sequentially from the loaded edge, producing ring-like undulations orthogonal to the direction of compression. 
In contrast, when the fabrics are less tightly constrained, they exhibit nearly synchronous emergence of one-sided helical (counterclockwise) wrinkles that span the entire sample, similar to \textit{spirality}.
The geometry of the resulting patterns is strongly linked to local loop kinematics, the mechanical response of the fabric, and the observed helical deformation modes, and is further modulated by minor asymmetries inherited from the knitting process. Our findings clarify how microscale loop arrangements govern global buckling modes in cylindrical knitted systems. The insights obtained here pave the way for a rational engineering strategy of knitted materials that display tunable mechanical behavior and programmable three-dimensional morphologies.

\begin{figure*}[ht!]
    \centering
    \includegraphics[width=1.0\textwidth]{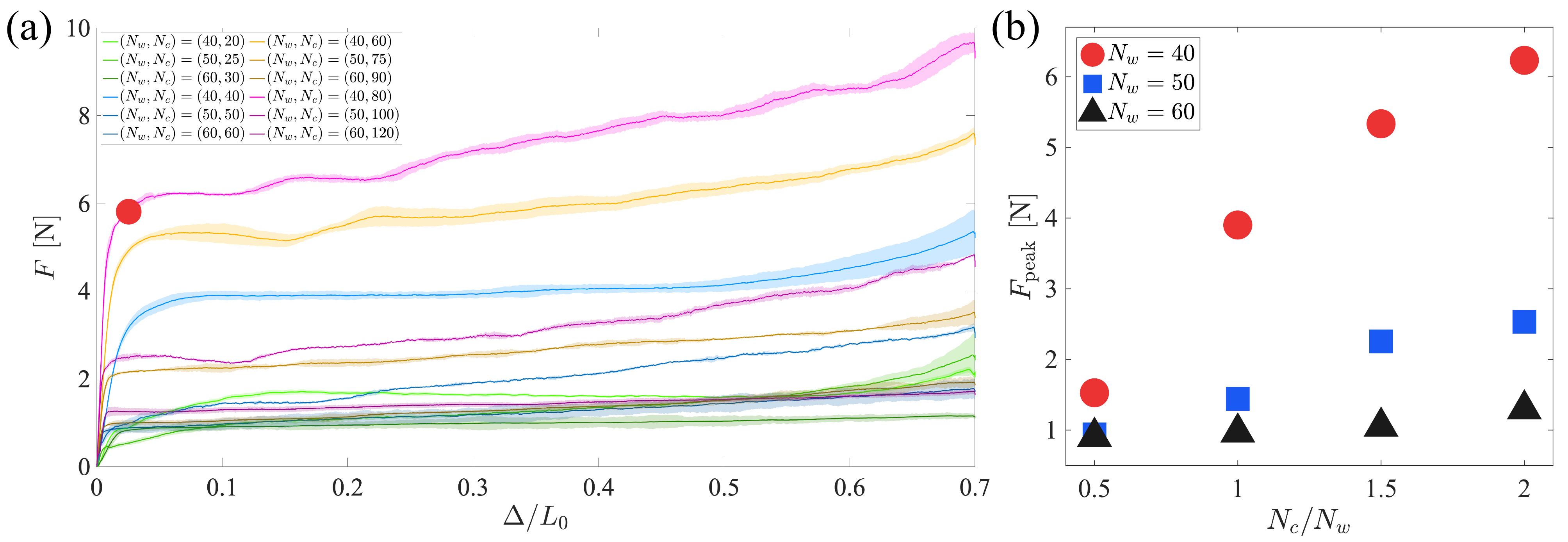}
    \caption{
    (a) Force–strain curves for samples for several sets of knitting numbers, $(N_w, N_c)$. 
    We classify the color of the curves based on the aspect ratio, $N_c/N_w$, and tightness, $N_w$; $N_c/N_w = 0.5, 1.0, 1.5$ and $2.0$ as green, blue, yellow, and magenta, respectively. Each darker color represents the result for a larger $N_w$.
    A representative peak force $F_{\rm peak}$ is highlighted by a filled circle on the curve. 
    (b) Relationship between the peak force $F_{\rm peak}$ obtained from the force–strain curves in (a) and the aspect ratio, $N_c⁄N_w$. 
    }
    \label{fig:2}
\end{figure*}

\section{Problem Definition}

The buckling behavior of cylindrical knitted fabrics wrapped around a rigid cylinder under axial compression is examined by preparing knitted tubes with controlled geometric parameters. The geometry of each sample is defined by two knitting numbers: the walewise (circumferential) knitting number $N_w$, specifying the number of loops around the tube, and the axial knitting number $N_c$, specifying the number of loops along the cylinder axis. The definitions of $N_w$ and $N_c$ are illustrated in Fig.~\ref{fig:1}(c) inset. 
Systematic variation of these knitting numbers enables an investigation of how loop-level geometry governs the onset and evolution of macroscopic wrinkling.

We wrap a cylindrical sample composed of plain-knit structures with the given set of knitting numbers, $(N_w, N_c)$, around a rigid cylinder. The cylinder diameter is fixed throughout, while the walewise knitting number, $N_w$, is varied to adjust the tightness of the knit.
The compressive load of the knit, $F$, is measured as a function of displacement, $\Delta$, using a uniaxial force-testing machine, and wrinkle formation is quantitatively analyzed from synchronized video recordings. This experimental framework enables the evaluation of the coupling between loop-scale geometry and macroscopic buckling modes.

\subsection{Fabrication of cylindrical knits}
\mbox{}\\
\indent
Cylindrical knitted samples are produced using a digital V-bed knitting machine (Kniterate, Spain) with 100 $\%$ cotton yarn (Emmy Grande, Olympus, Japan).
All samples adopt a plain-knit structure, the most fundamental architecture in knitted fabrics. The walewise knitting number is varied within the range $N_w = 40-60$, while the axial knitting number $N_c$ is adjusted to achieve knitting number ratios $N_c/N_w = 0.5-2$. This configuration allows us to prepare samples at different overall scales while maintaining identical knitting-number ratios.
The \textit{stitch size} is fixed at a value of $6$ across all samples, representing a dimensionless Kniterate parameter in which larger values produce looser loops. The knitting speed is maintained at $300$ mm/s, and the \textit{roller-advance} is set to $600$. The \textit{roller-advance} setting is specific to Kniterate and controls yarn tension during fabrication; larger values increase tension, thereby influencing residual stress in the knitted fabric.
After fabrication, all samples are placed in an acrylic container and vibrated to minimize external disturbances. They are then equilibrated under standard atmospheric conditions to ensure uniform relaxation prior to testing.

\subsection{Experimental methods}
\mbox{}\\
\indent
To characterize the buckling behavior of the knitted cylinders, each sample is mounted on a rigid cylinder with a diameter of $70$ mm and subjected to axial compression using a testing machine (EZ-LX, Shimadzu, Japan). Both ends of the knitted fabric are clamped between acrylic plates featuring custom-made comb-like protrusions. This clamping geometry ensures reliable engagement with the loops, preventing slippage and suppressing non-uniform deformation during the mechanical testing. The upper acrylic plate is attached to the crosshead of the testing machine, enabling controlled axial compression.
The gravitationally sagging state of each sample defines the zero-force condition $F = 0$, and the corresponding length is obtained as the natural length $L_0$. Compression tests are conducted at a crosshead speed of $v = 5$ mm/s, and the imposed displacement $\Delta$ is limited to 30 $\%$ of $L_0$. For each sample, three replicate compression tests are performed to confirm reproducibility. 
During compression, the force response associated with wrinkle formation is continuously recorded. Simultaneously, the deformation process is captured by a front-view camera to track loop geometry and wrinkle evolution. Representative snapshots of the compression process for different knitting number ratios are shown in Fig.~\ref{fig:1}(a)--(c). Quantitative measures of wrinkling behavior are subsequently extracted using image analysis (Image Processing Toolbox, MATLAB). 

\section{Buckling of cylindrical knit \& Loop deformation}

\subsection{Force-displacement curve of cylindrical knits}
\mbox{}\\
\indent
Typical snapshots from the compression tests are shown in Fig.~\ref{fig:1}, illustrating how knit buckling depends on the tightness around the cylinder. We vary the walewise knitting number, $N_w$, while keeping the cylinder diameter fixed to vary the tightness of the cylindrical knit. Samples with smaller $N_w$ (Fig.~\ref{fig:1}(a)(b)) exhibit a gradual emergence of vertical wrinkles originating from the loading side. In contrast, samples with larger $N_w$ develop spiral wrinkles that appear almost simultaneously across the entire specimen(Fig.~\ref{fig:1}(c)). These distinct deformation modes indicate that the loop architecture and the degree of circumferential constraint play a central role in governing the onset and evolution of surface instabilities under compression.

The corresponding force–strain responses are presented in Fig.~\ref{fig:2}(a). As observed in these curves, the compressive force $F$ shows a sharp initial rise followed by a more gradual increase. In this study, the point at which this steep rise first appears in the force–displacement curve is defined as the peak force $F_{\rm peak}$, which corresponds to the critical force for buckling, as highlighted by a marker in Fig.~\ref{fig:2}(a). 
A comparison of $F_{\rm peak}$ across samples with different geometric parameters is presented in Fig.~\ref{fig:2}(b). According to Fig.~\ref{fig:2}(b), samples with a smaller walewise knitting number $N_w$—corresponding to tighter wrapping around the cylindrical core—exhibit a more pronounced influence of the knitting number ratio. In particular, samples with a larger knitting number ratio and tighter wrapping exhibit higher values of $F_{\rm peak}$. This remarkable dependence of the peak force on the knit aspect ratio suggests that both increased yarn–yarn interactions and enlarged contact area with the cylindrical surface contribute to enhanced stiffness under external loading. 
Consequently, the response is dominated by local nonlinear deformations arising from the bending and frictional characteristics of the loop structure, and these effects manifest as the observed variations in $F_{\rm peak}$. Our findings imply that the local nonlinear deformation associated with loop bending and friction governs the mechanical response and that the observed variation in $F_{\rm peak}$ reflects the influence of compression-induced changes in loop geometry.

\subsection{Deformation of single loop structure}

\begin{figure}[t!]
    \centering
    \includegraphics[width=1.0\linewidth]{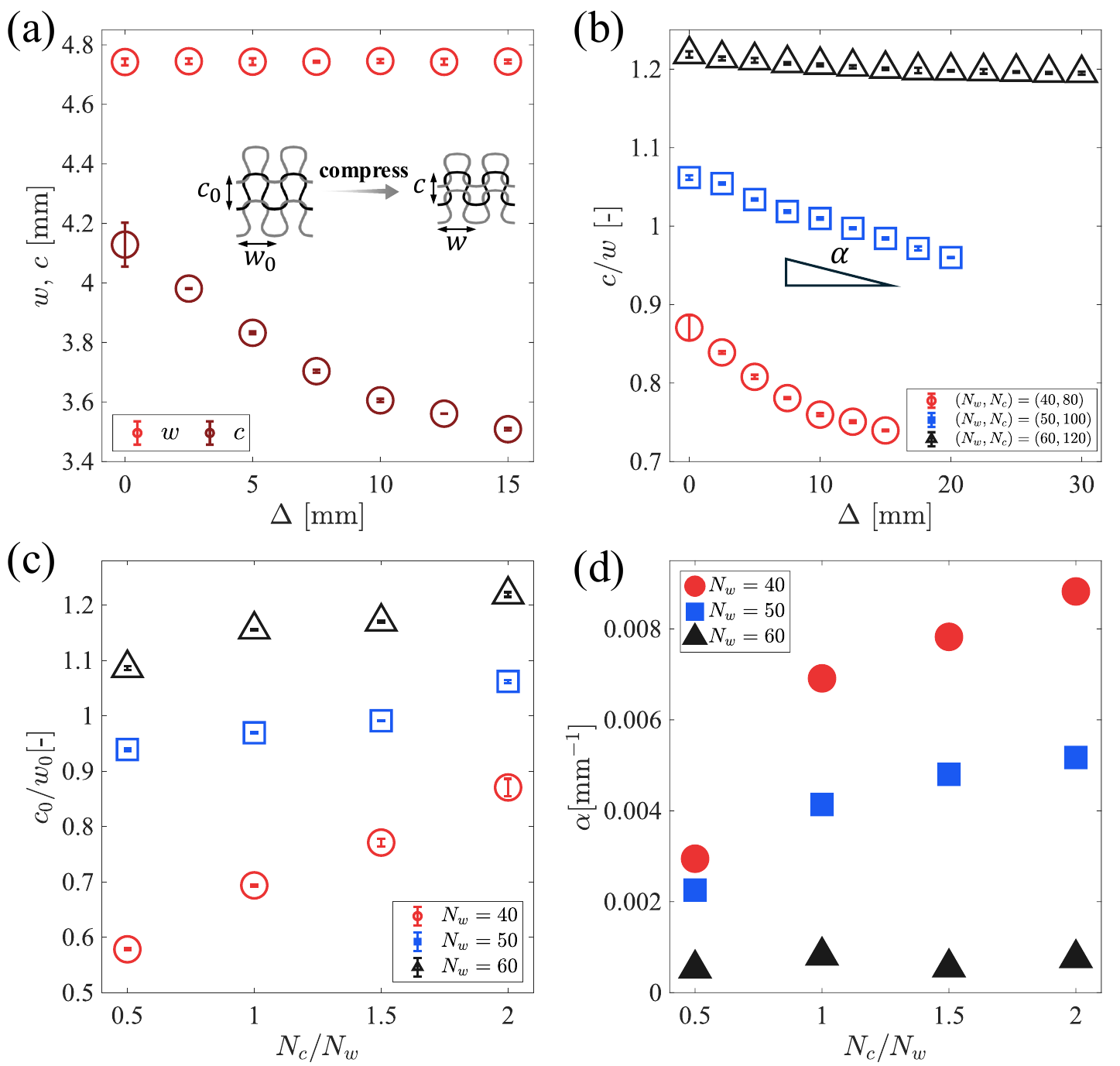}
    \caption{
    Characteristic length scales of the unit loop. 
    (a) The inset illustrates the initial $(w_0, c_0)$ and compressed $(w, c)$ wale and course lengths, and the plot shows $c$ and $w$ as functions of axial displacement $\Delta$ for a representative sample with $(N_w, N_c) = (40, 80)$.
    (b) The change of the aspect ratio of the loop, $c/w$ upon compression up to the buckling ($F\lesssim F_{\rm peak}$) for $(N_w, N_c) = (40, 80), (50, 100), (60, 120)$.
    (c) The ratio of the initial wale and course lengths, $c_0/w_0$, against the aspect ratio of the knitting number, $N_c/N_w$.
    (d) (Linear) Decrease ratio of $c/w$ upon axial compression, $\alpha$, plotted as a function of $N_c/N_w$ (See definition of $\alpha$ in (b)).
    }
    \label{fig:3}
\end{figure}

\mbox{}\\
\indent
The macroscopic mechanical responses described above, where the peak force, $F_{\rm peak}$, depends remarkably on the knitting numbers, $(N_w, N_c)$, indicate that the underlying loop-level geometry plays an essential role in the onset of buckling. To clarify this connection experimentally, we next examine how individual loop units deform prior to buckling instability and how these microscopic deformations correlate with the observed force–displacement behavior.
The geometry of a {``$\Omega$-shaped"} knitted loop is characterized using two geometric parameters defined in the inset of Fig.~\ref{fig:3}(a):
The wale length $w$, which represents the horizontal spacing between neighboring loops arranged side by side, and the course length $c$, which represents the vertical spacing between loops stacked in the knitting direction.
The corresponding initial dimensions prior to loading ($\Delta = 0$) are denoted by the initial wale length $w_0$ and the initial course length $c_0$.
To quantify how these loop dimensions evolve during compression, image analysis is performed on the captured video frames (image-processing toolbox, MATLAB). The videos (60 fps) are converted to frame data at 0.5 s intervals, followed by binarization, contour extraction of the loop boundaries, and curve fitting. These procedures provide a quantitative measure of loop-shape deformation.
A single vertical column of loops located at the front center on the compression side is extracted from the top ten consecutive loop rows for image analysis. This region is selected because substantial deformations consistently occur there across all samples, and its nearly planar appearance minimizes the influence of cylindrical curvature, thereby ensuring reliable comparison across samples.
The wale, $w$, and course length, $c$, are measured for each frame within the displacement before the buckling (when the knit remains almost cylindrical) to obtain a time series of loop-shape evolution during compression. To suppress local fluctuations, the loop geometry is represented by the average over the ten extracted loop rows.

Fig.~\ref{fig:3}(a) presents a representative example of how the loop parameters $w$ and $c$ evolve during compression for the sample with $(N_w, N_c) = (40, 80)$. The wale length $w$ remains nearly constant throughout loading, whereas deformation is governed primarily by the reduction in the course length $c$. In other words, the loop structures are vertically compressed while the remaining horizontal lengths are almost preserved.
This characteristic response is consistently observed across all samples. Given that the change in $c$ dominates while $w$ remains nearly invariant, the aspect ratio $c/w$ effectively captures the essential features of the single-loop shape deformation. As shown in Fig.~\ref{fig:3}(b), this aspect ratio decreases almost linearly with the applied displacement $\Delta$. 
To understand how the geometry at the onset of compression influences subsequent deformation, the relationship between the initial aspect ratio $c_0/w_0$ and the knitting number ratio $N_c/N_w$ is examined in Fig.~\ref{fig:3}(c). Samples with smaller walewise knitting number $N_w$ and smaller ratio $N_c/N_w$, i.e., samples wrapped more tightly around the cylinder, exhibit systematically smaller values of $c_0/w_0$. 
The correlation between the loop aspect ratio and the knit aspect ratio is consistent with our previous work on planar knits~\cite {Kotone2025}. Comparing Fig.~\ref{fig:2}(b) and Fig.~\ref{fig:3}(c), we find that the knit with small $c_0/w_0$ (initial loop is vertically compressed) requires large forces for the buckling, $F_{\rm peak}$, where the initial geometric configuration already encodes the differences in mechanical response among samples.
To describe the deformation of the loop geometry in a simplified expression, we assume that the aspect-ratio change follows the empirical relation $c/w = c_0/w_0 - \alpha\Delta$, where $\Delta$ is the imposed displacement and $\alpha$ corresponds to the rate of change. Using this definition, the coefficient $\alpha$ is extracted from the slope of the plots in Fig.~\ref{fig:3}(b) and the resulting values are plotted as a function of $N_c/N_w$ in Fig.~\ref{fig:3}(d).
Fig.~\ref{fig:3}(d) demonstrates that $\alpha$ varies in a well-defined manner with the knitting numbers: samples with smaller $N_w$, corresponding to a tighter wrap around the cylinder, exhibit more pronounced geometric deformation. Importantly, these results reveal trends analogous to those of the peak force $F_{\rm peak}$, where samples exhibiting larger changes in loop aspect ratio show higher peak forces~(Fig.~\ref{fig:2}(b)). These observations demonstrate a direct link between loop-geometry deformation and the macroscopic mechanical response, indicating that variations in loop shape—both initially and during compression—play a critical role in governing the peak force under axial loading.

\section{Wrinkle formation of cylindrical knit}

\begin{figure}[b!]
    \centering\includegraphics[width=1.0\linewidth]{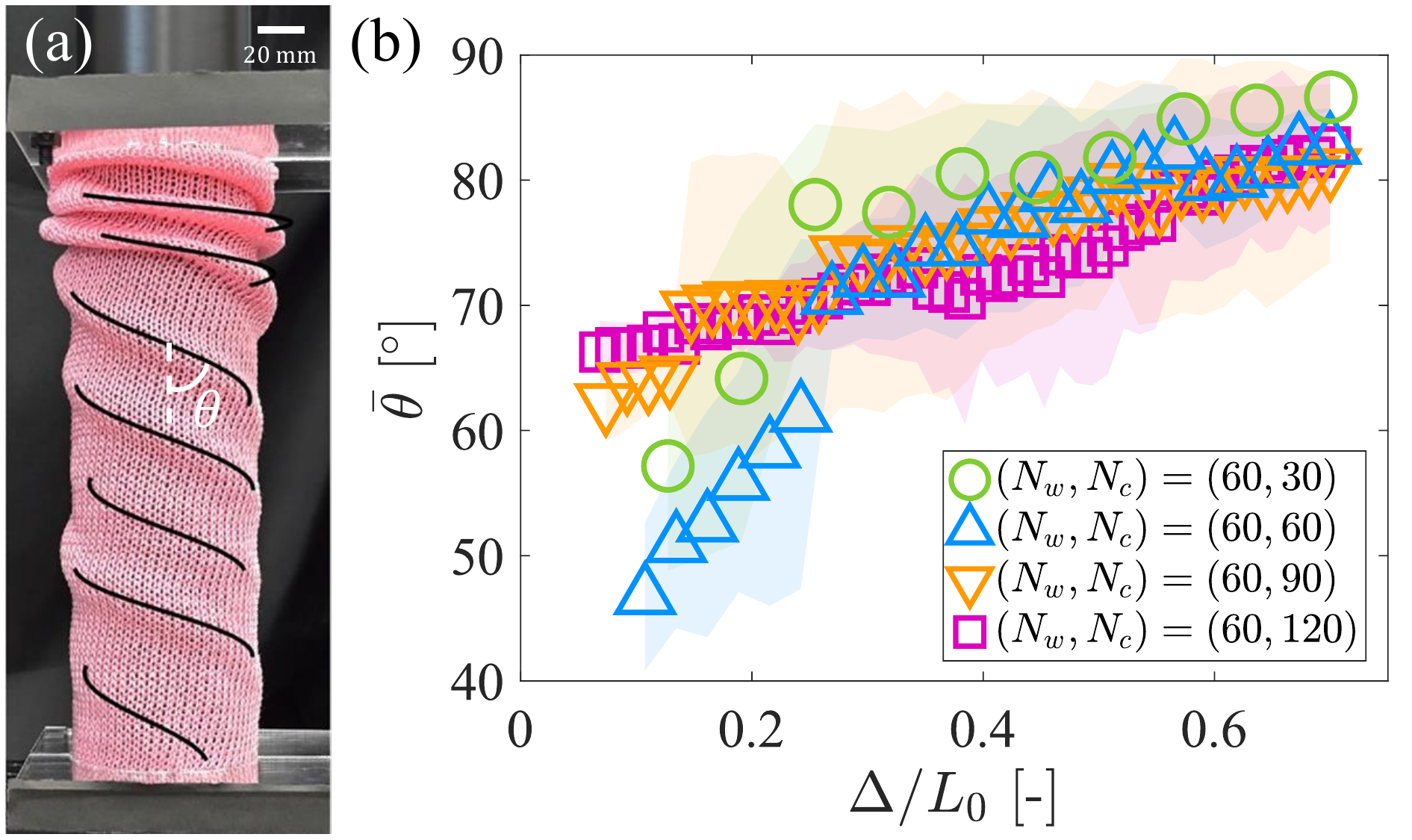}
    \caption{
    Helical buckling of cylindrical knits.
    (a) Typical example of image analysis used to extract wrinkle patterns and to define the pitch angle $\theta$ during compression of the knitted sample ($(N_w, N_c) = (60, 120)$). The extracted wrinkles approximated by sinusoidal curves are superimposed on the original images. The pitch angle for each wrinkle, $\theta$, is obtained from the tangent of the center of the corresponding sinusoidal curve.
    (b) 
    Variation of the average pitch angle $\bar{\theta}$ with the imposed displacement $\Delta$ for samples with $N_w = 60$. The average pitch angle, $\bar{\theta}$, is obtained by taking the average of $\theta$ over all wrinkles for a given $\Delta$ and the shaded regions represent the error bar. 
    }
    \label{fig:4}
\end{figure}

We have clarified the mechanical performance of cylindrical knits up to the onset of buckling. 
Here, we focus on the post-buckling behavior of cylindrical knits under axial compression. As described earlier, the wrinkle morphology depends strongly on the walewise knitting number $N_w$: samples with small $N_w$ (Fig.~\ref{fig:1}(a)(b)) develop wrinkles sequentially from the loading side, maintaining orientations nearly perpendicular to the compression direction, whereas samples with larger $N_w$ (Fig.~\ref{fig:1}(c)) exhibit simultaneously emerging helical wrinkles across the entire tube. 
We note again that a large $N_w$ corresponds to a loose-knit structure, because the cylinder diameter and the stitch size are fixed throughout. 
These contrasting observations indicate that loosely knitted samples favor a helical post-buckling mode, while tightly knitted samples transition into an accordion-like pattern, reflecting the dominance of frictional constraints when $N_w$ is small and the suppression of local rotational flexibility when $N_w$ is large.

\subsection{Pitch angles of helical buckling}
\mbox{}\\
\indent

To provide more quantitative evidence for the dependence of buckling behavior on tightness, we extract wrinkle geometry from video recordings of the compression tests. The recorded videos are converted into frame sequences at 0.5~s intervals, using the same procedure as for loop-scale shape measurements. In each frame, wrinkles are detected by exploiting the fact that surface undulations locally block incident light, forming strong brightness gradients. 
We extract the pixels of large intensity gradients and group them into each wrinkle. Assuming that the wrinkles are part of helices, we fit the set of pixel-locations by a sinusoidal curve, from which the helical pitch angle $\theta$ is computed (Fig.~\ref{fig:4}(a)). 
Here, $\theta$ is defined as the inclination of the sinusoidal trace of each wrinkle with respect to the cylinder axis. For example, the horizontally-developed wrinkle corresponds to $\theta = 90^{\circ}$. For each video frame, we compute the average of $\theta$ across all wrinkles, yielding the average pitch angle, $\bar{\theta}$, to characterize the average helical shape for a given displacement, $\Delta$. This process enables frame-by-frame quantification of wrinkle evolution during compression. 
Using this method, we analyze a representative sample with clearly observable helical deformation ($N_w = 60$).

The average pitch angle $\bar{\theta}$ for each frame is computed as a function of imposed compression $\Delta$ as Fig.~\ref{fig:4}(b). The pitch angle increases monotonically with compression, approaching $\approx90^\circ$ at large compression ratios, indicating that wrinkles gradually rotate to become nearly perpendicular to the loading direction as a more uniform post-buckling band forms. 
A comparison across samples with various knitting number ratios $N_c/N_w$ reveals that the samples with large $N_c/N_w$ ($= 1.5, 2.0$) exhibit a linear and stable increase of $\bar{\theta}$. In contrast, the pitch angles of $N_c/N_w = 0.5, 1.0$ increase immediately upon buckling and then follow the same trend as $N_c/N_w = 1.5, 2.0$. We anticipate that the stable increase in longer samples would arise from the increased inter-loop constraint, thereby leading to the collective propagation of helices.
These results demonstrate, with quantitative image-based measurements, that the macroscopic post-buckling behavior shifts from a loose, helical configuration in small $N_w$ samples to a tight, accordion-like mode in large $N_w$ samples.

\subsection{Mechanisms of helical buckling}
\mbox{}\\
\indent

Having clarified how the knitting geometry governs the emergence and evolution of helical wrinkles, we next examine factors that determine the direction of the helical twisting, which consistently appears counterclockwise in our experiments.
In axial compression experiments, helical wrinkles that emerge on knitted cylinders consistently spiral counterclockwise. Helical twisting in plain knitted fabrics, known as \textit{spirality}, typically originates from manufacturing-induced asymmetries such as yarn twist, knitting direction, and loop geometry \cite{Pavko2015}. These factors generate residual torque in the fabric, influencing both the direction and the intensity of the resulting twist. The counterclockwise helical wrinkles observed here are therefore likely rooted in such inherent asymmetries associated with the knitting direction.

\begin{figure}[t!]
    \centering\includegraphics[width=1.0\linewidth]{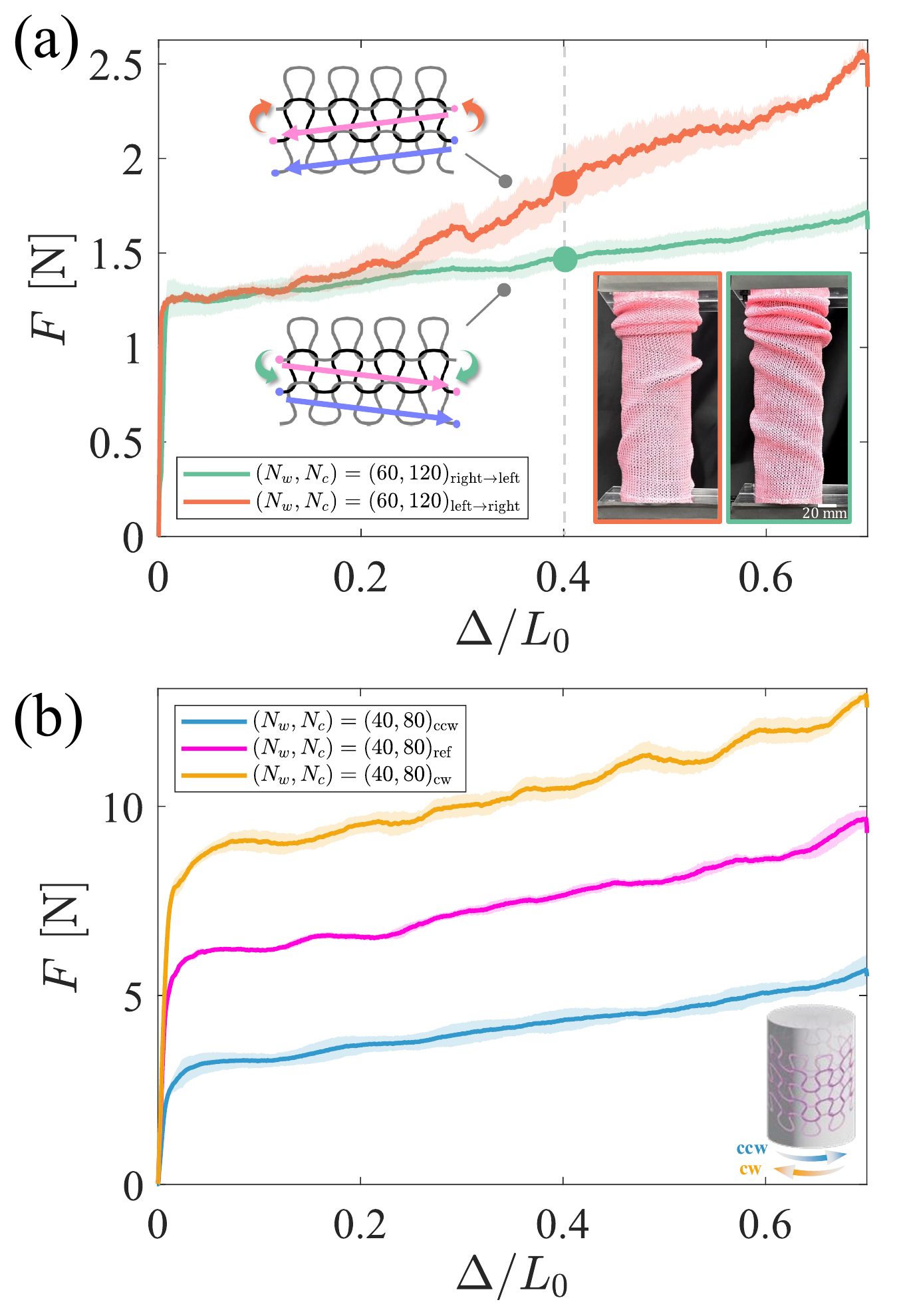}
    \caption{
    Roles of intrinsic (natural) twist of cylindrical knits.
    (a) The force responses of the sample with the same knitting numbers, $(N_w, N_c) = (60, 120)$, manufactured in the different knitting order. We plot the force displacement curve of the sample obtained through the knitting processes (right$\to$left) illustrated in the main text as a green curve (replotted from Fig.~\ref{fig:2}(a)). The orange curve represents the force response of the sample with the opposite knitting order. The snapshots of $\Delta/L_0 = 0.40$ are shown in the inset.
    (b) Force displacement curves of pre-twisted cylindrical knits $(N_w, N_c) = (40, 80)$. We compare the force-response of the knit pre-twisted by $\pm180^{\circ}$ (ccw, cw, respectively) with the twistless sample (denoted as ref). The result of the twistless case is replotted from Fig.~\ref{fig:2}.
    }
    \label{fig:5}
\end{figure}

The samples examined in this study are fabricated by connecting the two lateral edges of a rectangular knitted fabric on the knitting machine to form a cylindrical tube. We expect that this knitting operation constrains the loop topology at the seam and embeds structural asymmetry determined by the knitting direction and the orientation in which the loops are drawn and interlaced, with the loop-drawing direction in the initial knitting setup proceeding from right to left. 
To demonstrate that helical buckling is due to the manufacturing process, we perform mechanical testing on two samples knitted in opposite directions, with the knitting parameters fixed at $(N_w, N_c) = (60, 120)$. One sample is knitted as before, while the other is manufactured with the motorized knitting bead programmed to stroke in the opposite direction. The results of mechanical tests for the former and the latter are compared in Fig.~\ref{fig:5}(a) as right$\to$left and left$\to$right, respectively, where we find the remarkable difference in the mechanical performances.
This difference in mechanical testing is attributed to the contact geometry with the rigid cylinder (inner structure of the tube). In plain knitted fabrics, the front stitch corresponds to the surface where the $\Omega$-shaped loop head crosses above the legs, whereas the back stitch corresponds to the surface where the loop legs cross above the head. Although the loop structure is geometrically symmetric, the convention distinguishes the two surfaces based on the viewing direction, and the crossing order of yarn segments differs between the front and back stitches \cite{Amanatides2022}. 
Therefore, the reversal of the knitting is equivalent to flipping front and back stitches or to turning the cylindrical knit inside out, which alters the interacting surface with the cylinder between right$\to$left and left$\to$right in Fig.~\ref{fig:5}(a). The corresponding experimental snapshots (inset of Fig.~\ref{fig:5}(a)), where the handednesses of the wrinkles are inverted, highlight the fact that these samples are in an inside-out relationship. In summary, we have indirectly shown that the helical buckling of cylindrical knits originates from the intrinsic chirality embedded upon manufacturing.

\subsection{Buckling of pre-twisted cylindrical knits}
\mbox{}\\
\indent

We have shown that the cylindrical knits of plain-knitting have an intrinsic twist. Here, we demonstrate that the applied twist would enhance or suppress the axial stiffness of the cylindrical knit, which would also support our finding in the previous section. We perform axial compression against the knit of $(N_w, N_c) = (40, 80)$ with the $\pm180^{\circ}$ pre-twist by applying the rotation at the bottom end before compression with respect to the natural configuration. 

We reveal that samples pre-twisted in the counterclockwise direction, consistent with the intrinsically observed helical orientation, exhibit a smaller peak force in the force–strain curve, as shown in Fig.~\ref{fig:5}(b). Cylindrical knits buckle more readily when the imposed torsion aligns with the intrinsic asymmetry of the knitted structure. This result indicates that the buckling mode activated during compression is more readily triggered when it is compatible with the preexisting structural bias inherent to the loop topology.
The reproducible tunability of buckling strength against the pre-twist implies that the loop geometry and mechanics are highly entangled.
The wrinkle-formation process in cylindrical knits is governed not only by fabrication-induced asymmetries associated with the knitting direction but also by the three-dimensional loop geometry, which plays a crucial role in determining the mechanical interaction between the fabric and the cylindrical substrate. 
The observed counterclockwise helical wrinkles, therefore, do not arise accidentally but rather emerge from structural factors intrinsic to the knitting direction and seam topology, which collectively predispose the fabric toward a specific buckling pathway.

\section{Discussion and Summary}

In this study, we experimentally investigate the buckling and post-buckling behavior of plain knitted fabrics wrapped around a rigid cylinder under axial compression. By systematically varying the walewise and coursewise knitting numbers, we can reproducibly obtain wrinkle morphology and mechanical response of cylindrical knits, where the local loop structures are relevant. In particular, tightly wrapped knits with smaller walewise knitting numbers predominantly exhibit sequential, accordion-like wrinkles oriented perpendicular to the compression direction. In contrast, loosely wrapped knits favor the simultaneous emergence of helical wrinkles spanning the entire cylindrical surface. These distinct deformation modes are accompanied by systematic variations in the compressive force–displacement curves, the peak buckling force, and the evolution of loop geometry during compression.

To interpret our observations, it is instructive to discuss first the behavior of cylindrical knitted fabrics in the context of classical buckling phenomena~\cite{Koiter1945, vanderHeijden2009, Grabovsky2015}. A natural reference system is the diamond-shaped buckling patterns of axially compressed elastic cylindrical (continuous) shells. The expected buckling patterns of the elastic shell span across samples, both circumferential and axial directions~\cite{Grabovsky2015}, which are distinct from the wrinkle patterns of the cylindrical knits discussed in this paper. 
Despite the apparent similarity in global geometry and loading conditions, the wrinkle morphologies observed in knitted fabrics differ fundamentally from those of elastic shells. Elastic shells behave as continuous, approximately isotropic media, in which axial and circumferential strains are intrinsically coupled, leading to two-dimensional buckling patterns that minimize elastic energy~\cite{Rotter2006}. 
In contrast, the knitted fabrics studied here are discrete structures composed of interconnected $\Omega$-shaped loops~\cite{Kotone2025}. 
Under axial compression, a significant fraction of the imposed deformation can therefore be accommodated through yarn bending, torsion, and local reorganization at loop–loop contact points, rather than through in-plane stretching of a continuous elastic sheet. This discrete, kinematically rich structure provides a key distinction from classical shell behavior, particularly evident in helical buckling of cylindrical knits.

To further explore possible microscopic origins of the observed macroscopic behaviors, we briefly consider the three-dimensional geometry of individual knitted loops. A knitted loop is inherently a three-dimensional curved structure~\cite{Leaf1955, Kotone2025}, and its curvature distribution, contact area, and orientation can differ depending on whether the front or back stitch surface is in contact with the substrate. Such differences in exposed loop geometry may modify local contact conditions during compression. 
Establishing a direct and quantitative connection between loop-level features above buckling threshold and the measured force–displacement responses, however, lies beyond the scope of the present study and requires future investigations, for example, through numerical simulations that explicitly resolve yarn-scale mechanics and contact interactions.

In summary, our study establishes an experimental link between loop-scale geometry, macroscopic mechanical response, and wrinkle morphology in cylindrical knitted fabrics under axial compression. This experimental connection reveals the essential role of loop-level kinematics in governing global deformation in plain knitted fabrics under compression. At the same time, the effects of yarn-scale mechanics, friction, and loop asymmetry are treated only qualitatively. Addressing these factors through detailed numerical modeling, together with experiments on a wider range of stitch patterns, materials, and boundary conditions, is crucial for building a quantitative framework that connects loop-scale mechanics to macroscopic response. More broadly, beyond providing fundamental insight into instabilities in discrete textile structures, our results indicate that knitted fabrics constitute a simple and highly tunable platform for soft mechanical elements and actuators that exploit controlled wrinkling and chiral deformation~\cite{Jiahao2023, Ding2024, Ghorbani2024}.



\bibliographystyle{asmejour}   

\bibliography{asmejour-sample} 



\end{document}